# Classical and quantum mechanical plane switching in $CO_2$

Michaël SANREY and Marc JOYEUX[#],

*Laboratoire de Spectrométrie Physique (CNRS UMR 5588),*

*Université Joseph Fourier - Grenoble 1,*

*38402 St Martin d'Hères, France*

Dmitrii A. SADOVSKII[§]

*Université du Littoral, UMR 8101 du CNRS,*

*59140 Dunkerque, France*

[#] email : Marc.JOYEUX@ujf-grenoble.fr

[§] email : sadovski@univ-littoral.fr

**Abstract :** Classical plane switching takes place in systems with a pronounced 1:2 resonance, where the degree of freedom with lowest frequency is doubly-degenerate. Under appropriate conditions, one observes a periodic and abrupt precession of the plane in which the doubly-degenerate motion takes place. In this article, we show that quantum plane switching exists in $CO_2$ : Based on our analytical solutions of the classical Hamilton's equations of motion, we describe the dependence on vibrational angular momentum and energy of the frequency of switches and the plane switching angle. Using these results, we find optimal initial wave packet conditions for $CO_2$ and show, through quantum mechanical propagation, that such a wave packet indeed displays plane switching at energies around 10000 $cm^{-1}$ above the ground state on time scales of about 100 fs.

# 1 - Introduction

The resonant swing spring - a system which consists of a spring with one end fixed, a mass attached at the other end, and which is submitted to a constant vertical gravitation field - has recently received much attention from mathematicians and theoretical physicists [1-6]. When the physical parameters are chosen such that the stretching motion and the doubly-degenerate swinging motion are in 1:2 resonance, the swing spring has some remarkable features, including energy exchange, monodromy [3,5-8] and precession or switching of the swing plane [2-4]. By "precession of the swing plane" we mean the following. Suppose that the system is initially excited almost exclusively in the stretch degree of freedom. One observes that the system periodically evolves into a planar swinging motion before returning back to its original springing motion. Most interestingly, the orientation of the swinging plane changes by a constant amount from one swinging phase to the next one, the size of the step depending on the initial conditions [2-4]. Under appropriate conditions, the orientation of the plane remains nearly constant for a time of several pendular oscillations and then changes abruptly. Such behavior is called "plane switching".

While most earlier studies investigated the properties of the classical swing spring model, the aim of our present paper is to uncover this plane switching phenomenon in a real microscopic, i.e. quantum, system and to determine whether such a phenomenon can be observed experimentally. As pointed out in Ref. [5], the $CO_2$ molecule is a nearly perfect quantum analogue of the swing spring model. To analyze this real system, we should however take into account the detuning from the exact 1:2 bend:stretch Fermi resonance, anharmonicities and/or other degrees of freedom (antisymmetric stretch, rotation), and the last but not the least, the quantum nature of this system. The standard method to tackle this latter aspect of the problem relies on the study of non-stationary quantum states, i.e. wave packets, which in the limit of $\hbar \to 0$ evolve along the classical trajectories. More specifically, we are going to center a gaussian wave packet of minimal uncertainty $\hbar$ on the classical trajectory which exhibits the classical plane switching phenomenon. We want to find out whether and, especially, under which conditions this wave packet is capable of following the central trajectory for at least several steps and thus display quantum plane switching.

As in Ref. [5], we construct and use an integrable approximation, which describes explicitly only the three relevant vibrational degrees of freedom of $CO_2$, namely, the symmetric stretch (mode 1) and the doubly-degenerate bend (mode 2). We verify that this Hamiltonian is valid up to at least 10000 cm$^{-1}$ above the ground state, that is, 14 quanta of excitation in the bend. In order to understand the dependence of the phenomenon on the parameters of this system, the vibrational angular momentum $L$, the value of the action integral $I$ associated with the 1:2 polyad number, and the internal polyad energy, we solve the classical equations of motion for the 4th order Hamiltonian (Sect. 2). This solution relies on the early work [9,10] and extends Refs [2-4], where a 3rd order model without detuning was considered. We discuss classical plane switching in Sect. 3. Construction of optimal wave packets begins in Sect. 4. The average total energy of the wave packet is roughly determined by $P$. We choose this energy to be as high as possible (within the validity of our Hamiltonian) to maximize the density of states. At fixed $P$, the obtained expressions for the classical frequencies of the system and the swinging plane precession angle are then used to conjecture in Sect. 4 what the optimal values of $L$ and energy for the central trajectory are, in order for the plane switching phenomenon to show up as clearly as possible in the quantum mechanical system . By explicitly propagating the wave packet, we finally show in Sect. 5 that quantum $CO_2$ indeed displays plane switching.

While the main body of this article is hopefully of interest to the broad community of spectroscopists and chemists - and has consequently been written using adapted vocabulary and skipping a large part of the geometrical background - we feel that a number of aspects of Sect. 2-5 should be discussed in more details, in order to make this article also more accessible to the community of theoretical physicists and mathematicians and to relate it to preceding work in these fields. This more complete discussion is proposed in Appendix C.

**2 - Analytical resolution of Hamilton's equations**

The Hamiltonian of the integrable approximation used in Ref. [5] was obtained by applying 6th order Canonical Perturbation Theory [11] (CPT) to the potential energy surface of Zuniga et al [12]. Such an approximation reproduces experimental data to within a few wave numbers up to more than 10000 cm$^{-1}$ above the quantum mechanical ground state. In the present study, we stopped at 2nd order of CPT and obtained an Hamiltonian $H$, which is a polynomial of degree 4 in dimensionless normal coordinates. Of course, the agreement with experimental energies is consequently less quantitative, but the dynamics of the classical system is still correctly described qualitatively and, most importantly, Hamilton's equations can be solved analytically, while this is no longer possible at higher orders. On the other hand, numerical calculations in Sect. 4 and 5 can be performed with Hamiltonians of any order, but the differences with the results obtained using the 2nd order CPT Hamiltonian are so slight that we chose, for the sake of consistency, to present only these later ones throughout the paper. Note also that excitation of the uncoupled third mode of $CO_2$ (the antisymmetric stretch) essentially shifts all quantities upwards by the corresponding amount of energy. The results shown below were obtained by freezing this degree of freedom and setting the corresponding action integral to 1/2. The working Hamiltonian is thus written as

$$H = H_D + H_F$$
$$H_D = \omega_1 I_1 + \omega_2 I_2 + x_{11} I_1^2 + x_{12} I_1 I_2 + x_{22} I_2^2 + x_{LL} L^2 \quad (2.1)$$
$$H_F = 2k\sqrt{I_1(I_2^2 - L^2)} \cos(\phi_1 - 2\phi_2)$$

where $\omega_1$=1343.85 cm$^{-1}$, $\omega_2$=666.67 cm$^{-1}$, $x_{11}$=-2.88 cm$^{-1}$, $x_{12}$=-5.71 cm$^{-1}$, $x_{22}$=1.74 cm$^{-1}$, $x_{LL}$=-1.50 cm$^{-1}$, and $k$=26.69 cm$^{-1}$ (note that $\omega_1 \approx 2\omega_2$), $H_D$ is the so-called Dunham expansion, and $H_F$ is the lowest order coupling term, which describes the 1:2 Fermi resonance between the symmetric stretch (mode 1) and the doubly-degenerate bend (mode 2). The variables $(I_1, \phi_1)$ are the action-angle variables of the non-degenerate harmonic oscillator associated with the symmetric stretch, $(I_2, \phi_2)$ are the total action and conjugate angle of the doubly-degenerate harmonic oscillator associated with the bend, and $L$ is the bending vibrational angular momentum. Since the system is axially symmetric, $L$ is strictly conserved and $H$ does not depend on its conjugate angle. Relations between these quantities, the polar coordinates and momenta, and the dimensionless Cartesian normal modes coordinates and momenta $(p_1, q_1, p_{2x}, q_{2x}, p_{2y}, q_{2y})$ are detailed in Appendix A. Note that, as shown in the same Appendix, this Hamiltonian is a higher-order analogue of the one in Eq. (29a) of Dullin et al [3]. It is also similar to the one in Eq. (4) of Ref. [6].

The system with Hamiltonian in Eq. (2.1) is integrable. To show this, we rewrite $H$ in terms of the conjugate variables $(I, \phi_I)$ and $(J, \phi_J)$, where

$$\begin{aligned} I &= 2I_1 + I_2 \\ J &= 2I_1 \\ \phi_I &= \phi_2 \\ \phi_J &= \frac{\phi_1}{2} - \phi_2 \end{aligned} \quad (2.2)$$

One obtains

$$E \equiv H = \omega I + \varepsilon J + x_{II} I^2 + x_{IJ} IJ + x_{JJ} J^2 + x_{LL} L^2 + K\sqrt{J((I-J)^2 - L^2)} \cos(2\phi_J) \quad (2.3)$$

with trivial linear relationships between the parameters of Eq. (2.1) and those of Eq. (2.3). It can be seen that Eq. (2.3) depends neither on the angle conjugate to *I* nor, of course, on that conjugate to *L*. Therefore, both *I* and *L* are constants of the motion and the system is integrable.

Recall that we want to describe analytically the period and the angle of the plane switching as functions of dynamical constants *L, I* and *E*. To that end, we consider the Hamiltonian in Eq. (2.3) as the Hamiltonian of the one degree of freedom reduced system with conjugate coordinates $(J, \phi_J)$. We note that $J = 2I_1$ is non-negative and bound from above by *I*. Therefore, *J* is a periodic function. In fact, *J* describes the energy in the stretch degree of freedom; the period *T* of $J(t)$ is the period of energy transfer between the two modes and, consequently, the period of plane switching. The position of the swinging plane is described by the angle conjugate to *L* in Eqs. (2.1) and (2.3), as has been demonstrated in Ref. [3]. It is emphasized that this angle, which is denoted by ρ in the rest of this article, is NOT the polar angle $\tan^{-1}(q_{2y}/q_{2x})$. The expression for ρ is provided in Appendix A. Simple calculations show that this definition is identical to the one in Eq. (54) of Ref. [3]. The angle of plane switching can be defined as $\Delta\rho = \int_t^{t+T} \frac{d\rho}{dt} dt$. Note that, as function of parameters *L, I* and *E*, Δρ is defined modulo π and special arrangements may be needed if one wants to define it continuously at certain given values of these parameters (see below).

Hamilton's equations of motions are solved along the same lines as in Refs. [9,10]. Using the Hamiltonian in Eq. (2.3), we obtain the equation for $dJ/dt$ in the form

$$\begin{aligned}(\frac{dJ}{dt})^2 &= (-\frac{\partial H}{\partial \phi_J})^2 \\ &= 4K^2 J((I-J)^2 - L^2) - 4(-E + \omega I + \varepsilon J + x_{II}I^2 + x_{IJ}IJ + x_{JJ}J^2 + x_{LL}L^2)^2 \\ &= -4x_{JJ}^2 (J-\alpha)(J-\beta)(J-\gamma)(J-\delta)\end{aligned} \quad (2.4)$$

where the roots α, β, γ and δ depend on *E*, *I* and *L* and satisfy the inequalities

$$0 \leq \delta \leq J \leq \alpha \leq \beta \leq \gamma$$
$$\alpha \leq I - L \tag{2.5}$$

Eq. (2.4) corresponds to Eq. (39) of Ref. [3]. Since we took first anharmonicities into account, the polynomial in the right-hand side of Eq. (2.4) is of degree 4 with respect to *J*, while it is of degree 3 in Ref. [3]. Moreover, Dullin et al consider only the case with no detuning ($\varepsilon = 0$). The solution of Eq. (2.4) is

$$J(t) = \beta + \frac{\alpha - \beta}{1 - \eta \operatorname{sn}^2(\lambda t | \mu)} \tag{2.6}$$

where sn() is the Jacobian elliptic function [13] and

$$\eta = \frac{\alpha - \delta}{\beta - \delta}$$
$$\mu = \eta \frac{\beta - \gamma}{\alpha - \gamma} \tag{2.7}$$
$$\lambda = \frac{1}{2}\sqrt{-4x_{JJ}^2(\alpha - \gamma)(\beta - \delta)}$$

To simplify the choice of the integration constant, we suppose in Eq. (2.6) that *J* is at maximum ($J = \alpha$) at time $t = 0$. As already stated above, *J*(*t*) is periodic. From Eq. (2.6), its period *T* is

$$T = \frac{2K(\mu)}{\lambda} \tag{2.8}$$

and its angular frequency

$$\omega^* = \frac{2\pi}{T} = \frac{\pi \lambda}{K(\mu)}, \tag{2.9}$$

where $K(\mu)$ is the complete elliptic integral of the first kind [13]. Eq. (2.6) can now be used to find ρ(*t*). Using Eq. (2.3), Hamilton's equation for ρ is recast in the form

$$\frac{d\rho}{dt} = \frac{\partial H}{\partial L} = (2x_{LL} + x_{JJ})L - \frac{1}{2}\frac{E - E_{I-L}}{I - L - J} + \frac{1}{2}\frac{E - E_{I+L}}{I + L - J} \tag{2.10}$$

where the two energies

$$E_{I\mp L} = \omega I + \varepsilon(I \mp L) + x_{II}I^2 + x_{IJ}I(I \mp L) + x_{JJ}(I \mp L)^2 + x_{LL}L^2 \tag{2.11}$$

are obtained by setting $J = I \mp L$ in Eq. (2.3). The solution of Eq. (2.10),

$$\rho(t) = r_0 t + r_1 \Pi(\xi; \lambda t|\mu) + r_2 \Pi(\zeta; \lambda t|\mu) , \tag{2.12}$$

is written in terms of elliptic integrals of the third kind [13], where

$$\begin{aligned}
r_0 &= (2x_{LL} + x_{JJ})L - \frac{1}{2}\frac{E - E_{I-L}}{I - L - \beta} + \frac{1}{2}\frac{E - E_{I+L}}{I + L - \beta} \\
r_1 &= -\frac{1}{2\lambda}(E - E_{I-L})(\frac{1}{I - L - \alpha} - \frac{1}{I - L - \beta}) \\
r_2 &= \frac{1}{2\lambda}(E - E_{I+L})(\frac{1}{I + L - \alpha} - \frac{1}{I + L - \beta}) \\
\xi &= \eta\frac{I - L - \beta}{I - L - \alpha} \\
\zeta &= \eta\frac{I + L - \beta}{I + L - \alpha}
\end{aligned} \tag{2.13}$$

Eq. (2.12) can be expressed in the form

$$\begin{aligned}
\rho(t) = &\left\{r_0 + \frac{2r_1}{T}\Pi(\xi|\mu) + \frac{2r_2}{T}\Pi(\zeta|\mu)\right\}t + \\
&r_1\left\{\Pi(\xi;\lambda\tau|\mu) - \frac{2\tau}{T}\Pi(\xi|\mu)\right\} + r_2\left\{\Pi(\zeta;\lambda\tau|\mu) - \frac{2\tau}{T}\Pi(\zeta|\mu)\right\}
\end{aligned} \tag{2.14}$$

where $\tau = \text{mod}(t, T)$, which clearly shows that $\rho(t)$ is the sum of a linear contribution (first term) plus a periodic contribution of period $T$ (second and third terms). The angle of plane switching is therefore obtained as $\Delta\rho = \rho(T) - \rho(0)$, that is

$$\Delta\rho = r_0 T + 2r_1\Pi(\xi|\mu) + 2r_2\Pi(\zeta|\mu) . \tag{2.15}$$

It is important to realize that the $\phi_J$ dependence was removed from Eqs. (2.4) and (2.10). Therefore, potential problems related to the ill-definition of $\phi_J$ at the $J = I$ and $L = 0$ singularity of Eqs. (2.12)-(2.14), which corresponds to the unstable relative equilibrium discussed later in Sect. 3, are circumvented. Approximations for $J$, $\omega^*$, $\rho$, and $\Delta\rho$ can even be obtained in the neighborhood of this singularity, by expanding the roots α and β for energies $E$ close to $E_{I-|L|}$ and replacing the Jacobi sn() function in Eq. (2.6) by the hyperbolic tangent function in the time interval $[-T/2, T/2]$. It can be shown in this way that, close to the singularity and for vanishing detuning and anharmonicities, $\Delta\rho$ has the form

$$\Delta\rho = -\text{sgn}(E - E_{I-|L|}) \left\{ \tan^{-1} \sqrt{\frac{u+1}{u-1}} - \tan^{-1} \sqrt{\frac{u-1}{u+1}} \right\} \quad (2.16)$$

where

$$u = \sqrt{1 + \frac{(E - E_{I-|L|})^2}{K^2 IL^2}} \quad (2.17)$$

This asymptotic expression is equivalent to the one in Eq. (72) of Ref. [3] (see also Ref. [14]). Finally, note that in the case without Fermi coupling ($k = K = 0$), $\rho(t)$ equals $2x_{LL}Lt$. The corresponding trivial contribution $2x_{LL}LT$ will be subtracted from $\Delta\rho$ in Sect. 3.

**3 - Classical plane switching in $CO_2$**

The expressions derived in Sect. 2 are singular at $J = I$ and $L = 0$. Since these conditions imply that $p_{2x} = p_{2y} = q_{2x} = q_{2y} = 0$ and $E = E_I$, the singularity takes place when all energy is put in mode 1, that is, for $CO_2$, in the symmetric stretch degree of freedom. In the original system, it corresponds to the unstable short periodic orbit, which is at the origin of monodromy and plane switching [3,5]. The singularity corresponds to an unstable equilibrium of the reduced one degree of freedom system with dynamical variables $(J, \phi_J)$ and Hamiltonian in Eq. (2.3), where $L$ and $I$ are considered as parameters. Classically [2-4], plane switching manifests itself most clearly for trajectories approaching closely this short periodic orbit, together with its stable and unstable manifolds, i.e. for trajectories with $L$ close to 0 and energy close to $E_I$. In consequence, this section presents a qualitative description of what happens for such trajectories.

When $L = 0$, the classical frequency $\omega^*$ goes logarithmically to zero when the energy approaches that of the unstable equilibrium [10], as illustrated in Fig. 1, which displays the variation of $\omega^*$ as a function of energy $E$ for a fixed value of $I$ and several values of $L$. In contrast, the period does not diverge when $L \neq 0$. Furthermore, variations of $\omega^*$ become smoother and smoother with increasing values of $|L|$.

The angle of plane switching $\Delta\rho - 2x_{LL}LT$ as function of $E$, $I$ and $L$ can be analyzed throughout the allowed energy range in a similar way, see Fig. 2. Since $\Delta\rho$ is an odd function of $L$, $\Delta\rho - 2x_{LL}LT$ remains zero for $L = 0$. For $L \neq 0$, $\Delta\rho$ in Eq. (2.15) jumps by $\pi$ at the two specific points of the parameter space for which $E = E_{I-|L|}$. As we said above, this is insignificant to our analysis because $\Delta\rho$ is modulo $\pi$ anyway, and it can be made artificially continuous by adding $\pi$ to the value given by Eq. (2.15) in the case $E > E_{I-|L|}$. However, the discontinuity reflects the change of $d\rho/dt$, which is positive when $E < E_{I-|L|}$ and negative when $E > E_{I-|L|}$. As in the case of $\omega^*$, $\Delta\rho$ changes rapidly in the neighborhood of $E = E_{I-|L|}$ for very small fixed values of $|L|$, while its evolution with $E$ becomes slower with increasing values of $|L|$.

The time evolution of $\rho$ for different values of $L$ and $E$ (see Fig. 3 and compare to Fig. 7 of Ref. [1] and Fig. 5 of Ref. [3]) can be understood easily on the basis of the above considerations. If $|L|$ is very small (top plot) and $E$ very close to $E_{I-|L|}$, then $\rho(t)$ looks like a stair, with sharp transitions of amplitude $\Delta\rho \approx \pi/2$ connecting very long plateaux ($T$ diverges at the relative equilibrium). According to Eqs. (2.10) and (2.12), the transitions occur when $J$ is near its maximum ($J \approx \alpha \approx I - |L|$), that is, when most of the energy is localized in the symmetric stretch degree of freedom. Each plateau corresponds to the oscillations of the bending motion near an almost fixed direction in the ($q_{2x}, q_{2y}$) plane, while the sharp transition between two successive plateaux represents the periodic abrupt change of this direction. It is precisely this behavior which is called plane switching [1-4].

If $|L|$ is kept small while $E$ departs from $E_{I-|L|}$, the amplitude of the steps diminishes rapidly, while $T$ is simultaneously reduced (see upper plot of Fig. 3). When $|L|$ becomes larger (middle and bottom plots of Fig. 3), the length of the plateaux substantially decreases, but ρ(t) evolves less markedly with energy. For example, it is seen that for $L = 2$ and $L = 6$ the two curves at $E = E_{I-L} + 0.1$ cm$^{-1}$ and $E = E_{I-L} + 40$ cm$^{-1}$ look like stairs of amplitude $\Delta\rho \approx \pi/2$ and are almost superimposed. Whatever the value of $L$, the stair-like character of ρ(t) disappears when energy deviates too much from $E_{I-|L|}$.

It should also be emphasized that, like for the swing spring, the bending motion appears to be "linear" only for very small $|L|$. Rigorously speaking, this motion traces ellipses in the ($q_{2x}, q_{2y}$) plane [3]. When $|L|$ is larger, the eccentricity of the ellipses becomes smaller, so that the bending motion can hardly be described as taking place along a line. Consider, for example, the ($q_{2x}, q_{2y}$) plane projections of trajectories with $I = 13.5$, $L$=0.1, 2 and 6, and $E = E_{I-L} + 0.1$ cm$^{-1}$ shown in Fig. 4. One is tempted to conclude, somewhat subjectively, that plane-switching, that is, motion along a line which precesses by jumps, takes place for $L = 0.1$ and $L = 2$ (and thus for any $L$ in the interval), but not for $L = 6$.

**4 - Determination of optimal initial conditions for the plane switching wave packet**

One of the principal objectives of this work is to determine the best initial conditions for a quantum wave packet to display plane switching. First we chose the total action *I* for the centre of the packet in order to have a sufficiently large density of states while remaining in the validity domain of our Hamiltonian. According to this criterion, values around *I*=14 (corresponding to energies near 10000 cm$^{-1}$ above the quantum mechanical ground state) appear reasonable. Indeed, we have seen in Ref. [5] that the related phenomenon of quantum monodromy can be well observed in this region.

It remains to determine the values of *E* and *L*. Recall from the previous section that the plane switching phenomenon is most clearly seen for trajectories launched in the vicinity of the singularity at $J = I$ and $L = 0$. Consequently, we are tempted to center our wave packets on such trajectories. At the same time, ω* and Δρ depend very sharply on the initial conditions in this region. Therefore, wave packets launched too close to the unstable relative equilibrium will disperse very rapidly, i.e. in a time less than *T*. If such rapid dispersion takes place, quantum plane switching cannot be observed. In other words, the rate of change of ω* and Δρ should not become too large on the scale of $\hbar$ ($\hbar = 1$ in the units of this paper). Examination of Fig. 3 suggests that the solution might consist in centering the wave packet at somewhat larger values of *L* and the corresponding energy $E_{I-L}$. The good compromise for $CO_2$ in the 10000 cm$^{-1}$ energy range therefore appears to be a wave packet centered around $L = 2$ and $E = E_{I-L}$.

We checked the validity of this conjecture by launching swarms of 10000 classical trajectories with a gaussian distribution of width $\Delta p_j = \Delta q_j = \frac{1}{\sqrt{2}}$ along each degree of freedom (*j*=1, 2x and 2y). We propagated the trajectories numerically using the Hamiltonian expressed in dimensionless normal coordinates (Eqs. (2.1) and (A.4) in Sect. 2 and Appendix A). As expected, we found that a swarm centered on a trajectory with $I = 13.5$, $L = 0.1$ and $E = E_{I-L}$, dispersed so rapidly, that the trajectories were almost uniformly spread over all the accessible phase space in times shorter than the period $T$ of the central trajectory. No plane switching was observed. The result for a distribution centered on a trajectory with $I = 13.5$, $L = 2$ and $E = E_{I-L}$, which is shown in Fig. 5, is qualitatively different. Each vignette of this figure shows the projection of the trajectories on the $(q_{2x}, q_{2y})$ plane at given increasing times $t$ indicated in the upper left corner (in fs). The position of the trajectory, around which the distribution was centered at $t = 0$, is indicated by a large empty circle. The first three vignettes ($t$=2, 26 and 50 fs) correspond to the first round trip along the initial direction of the bending motion. One observes little dispersion during this phase, because it corresponds to the plateau of $\rho(t)$ for most trajectories. Then, plane switching occurs near $t$=50 fs, and the last five vignettes ($t$=66, 90, 102, 114 and 126 fs) correspond to one round trip along a direction almost exactly perpendicular to the initial one. Dispersion is much more important during this second phase, because, as discussed above, the plane switching amplitude and the exact time where the step occurs are different for each trajectory. Dispersion in $(q_{2x}, q_{2y})$ is maximum at the "turning points" of the bending motion (i.e. the extremities of the narrow ellipses). Nonetheless, one clearly observes that the density of trajectories near the "central" trajectory is at all times significantly larger than elsewhere. Moreover, both the time required for the first plane switch and the switch angle correspond well to those of the "central" trajectory. This confirms that $L = 2$ and $E = E_{I-L}$ are likely to be suitable initial conditions to observe quantum plane switching in $CO_2$.

**5 - Quantum mechanical plane switching in $CO_2$**

To investigate quantum plane switching in $CO_2$, we considered a minimum uncertainty wave packet centered at time $t = 0$ on the same classical trajectory as in Sect. 4. We then computed its time evolution quantum mechanically. More precisely, the wave packet at time $t = 0$ is taken in the form

$$\Phi_{\overline{\mathbf{p}},\overline{\mathbf{q}}}(q_1,q_2,\varphi,t=0) = \Phi_{\overline{p}_1,\overline{q}_1}(q_1)\Phi_{\overline{p}_{2x},\overline{q}_{2x}}(q_2,\varphi)\Phi_{\overline{p}_{2y},\overline{q}_{2y}}(q_2,\varphi) \qquad (4.1)$$

where

$$\begin{aligned}\Phi_{\overline{p}_1,\overline{q}_1}(q_1) &= \pi^{-1/4}\exp\left\{i\overline{p}_1 q_1 - \frac{1}{2}(q_1-\overline{q}_1)^2\right\}\\ \Phi_{\overline{p}_{2x},\overline{q}_{2x}}(q_2,\varphi) &= \pi^{-1/4}\exp\left\{i\overline{p}_{2x} q_2\cos\varphi - \frac{1}{2}(q_2\cos\varphi-\overline{q}_{2x})^2\right\} \\ \Phi_{\overline{p}_{2y},\overline{q}_{2y}}(q_2,\varphi) &= \pi^{-1/4}\exp\left\{i\overline{p}_{2y} q_2\sin\varphi - \frac{1}{2}(q_2\sin\varphi-\overline{q}_{2y})^2\right\}\end{aligned} \qquad (4.2)$$

and $\overline{\mathbf{p}} = (\overline{p}_1,\overline{p}_{2x},\overline{p}_{2y})$ and $\overline{\mathbf{q}} = (\overline{q}_1,\overline{q}_{2x},\overline{q}_{2y})$ are the dimensionless normal coordinates of the central trajectory at time $t = 0$. We projected $\Phi_{\overline{\mathbf{p}},\overline{\mathbf{q}}}(q_1,q_2,\varphi,t=0)$ on the eigenvectors $\Psi_n(q_1,q_2,\varphi)$ of the quantum mechanical counterpart of the classical Hamiltonian in Eq. (2.1) (see Appendix B), leading to

$$\Phi_{\overline{\mathbf{p}},\overline{\mathbf{q}}}(q_1,q_2,\varphi,t=0) = \sum_j c_j \Psi_j(q_1,q_2,\varphi) \qquad (4.3)$$

where

$$c_j = \left\langle \Psi_j \middle| \Phi_{\overline{\mathbf{p}},\overline{\mathbf{q}}} \right\rangle_{t=0} = \int_0^{2\pi} d\varphi \int_0^{+\infty} q_2 dq_2 \int_{-\infty}^{+\infty} dq_1 \Psi_j^*(q_1,q_2,\varphi) \Phi_{\overline{\mathbf{p}},\overline{\mathbf{q}}}(q_1,q_2,\varphi,t=0) \qquad (4.4)$$

The leading contributions to the quantum wave packets are naturally due to the eigenfunctions, whose quantum numbers $P$ and $\ell$ (see Appendix B) are close to the values of their classical counterparts $I+2$ and $L$ for the central trajectory. The wave packet at time $t$ was then obtained in a standard way from

$$\Phi_{\overline{\mathbf{p}},\overline{\mathbf{q}}}(q_1,q_2,\varphi,t) = \sum_j c_j \Psi_j(q_1,q_2,\varphi)\exp(-iE_j t) \qquad (4.5)$$

where $E_j = \langle \Psi_j | H | \Psi_j \rangle$ is the eigenvalue associated with $\Psi_j$. To visualize plane switching, we plotted the probability density of the wave packet in the $(q_{2x}, q_{2y})$ plane

$$P(q_{2x}, q_{2y}, t) = \int_{-\infty}^{+\infty} dq_1 \left| \Phi_{\overline{\mathbf{p}}, \overline{\mathbf{q}}}(q_1, q_2, \varphi, t) \right|^2 \qquad (4.6)$$

The computed probability density is shown in Fig. 6 for the same time sequence as in Fig. 5. Apart from the sharper localization of the quantum wave packet at certain times (especially at $t$=50, 102 and 126 fs), the agreement between classical and quantum mechanical results is striking. Thus we can conclude that the density of states is sufficiently high and we are close to the semiclassical limit. Most importantly, the propagation of the quantum wave packet displays clearly the abrupt change of the orientation of the swinging plane.

## 6 - Conclusion

In this work, we demonstrated that quantum plane switching exists in $CO_2$. This quantum system displays, under appropriate conditions, a periodic and abrupt precession of the plane in which the doubly-degenerate bending motion takes place. Based on our analytical solutions of the classical equations of motion for the 4th order Hamiltonian, we described the dependence of the frequency of switches ω* and the plane switching angle Δρ on vibrational angular momentum and energy. Using these results, we found optimal initial wave packet conditions and showed, through quantum mechanical propagation, that such a wave packet indeed displays plane switching at energies around 10000 cm$^{-1}$ above the ground state on time scales of about 100 fs. This provides concrete motivation and vital information for further analysis of the possibility of experimental observation of plane switching in this system.

**Appendix A : Relations between the various sets of coordinates**

The action-angle variables $(I_1, \phi_1)$ of the non-degenerate harmonic oscillator associated with the symmetric stretch are related to the set of dimensionless normal coordinates $(p_1, q_1)$ through

$$q_1 = \sqrt{2I_1} \cos \phi_1$$
$$p_1 = -\sqrt{2I_1} \sin \phi_1 \qquad (A.1)$$

In Eq. (2.1), the doubly-degenerate bend is described by the two sets of conjugate variables $(I_2, \phi_2)$ and $(L, \rho)$, where $I_2$ is the total action, $\phi_2$ its conjugate angle, $L$ the bending vibrational angular momentum, and $\rho$ the angle conjugate to $L$ (remember that $\rho$ is the angle which describes the position of the "swinging plane"). Two transformations are needed to relate these coordinates to the $(p_{2x}, q_{2x})$ and $(p_{2y}, q_{2y})$ dimensionless normal ones, via $(p_2, q_2)$ and $(L, \varphi)$. These transformations write

$$q_2 = \sqrt{I_2 + \sqrt{I_2^2 - L^2} \cos(2\phi_2)}$$
$$p_2 = -\frac{1}{q_2}\sqrt{I_2^2 - L^2} \sin(2\phi_2) \qquad (A.2)$$
$$\varphi = \rho + \frac{1}{2}\arctan\left(\frac{L p_2 q_2}{L^2 - I_2 q_2^2}\right)$$

and

$$p_{2x} = p_2 \cos \varphi - \frac{L}{q_2} \sin \varphi$$
$$q_{2x} = q_2 \cos \varphi$$
$$p_{2y} = p_2 \sin \varphi + \frac{L}{q_2} \cos \varphi \qquad (A.3)$$
$$q_{2y} = q_2 \sin \varphi$$

The fact that the Hamiltonian of Eq. (2.1) is indeed a 4th order polynomial in terms of the dimensionless normal coordinates follows from the relations :

$$I_1 = \frac{1}{2}(p_1^2 + q_1^2)$$

$$I_2 = \frac{1}{2}(p_{2x}^2 + p_{2y}^2 + q_{2x}^2 + q_{2y}^2)$$

$$L = p_{2y}q_{2x} - p_{2x}q_{2y}$$

$$2\sqrt{I_1(I_2^2 - L^2)}\cos(\phi_1 - 2\phi_2) = \frac{1}{\sqrt{2}}(q_1(q_{2x}^2 + q_{2y}^2 - p_{2x}^2 - p_{2y}^2) + 2p_1(p_{2x}q_{2x} + p_{2y}q_{2y}))$$

(A.4)

**Appendix B : The quantum mechanical Hamiltonian**

The non-zero matrix elements of the quantum mechanical counterpart of the Hamiltonian of Eq. (2.1) in the direct product basis $|n_1, n_2, \ell\rangle = |n_1\rangle \otimes |n_2, \ell\rangle$ of the 1-dimensional and 2-dimensional harmonic oscillators are

$$\langle n_1, n_2, \ell | H | n_1, n_2, \ell \rangle = \omega_1(n_1 + \frac{1}{2}) + \omega_2(n_2 + 1) +$$
$$x_{11}(n_1 + \frac{1}{2})^2 + x_{12}(n_1 + \frac{1}{2})(n_2 + 1) + x_{22}(n_2 + 1)^2 + x_{LL}\ell^2 \quad \text{(B.1)}$$

$$\langle n_1, n_2, \ell | H | n_1 - 1, n_2 + 2, \ell \rangle = \langle n_1 - 1, n_2 + 2, \ell | H | n_1, n_2, \ell \rangle = -k\sqrt{n_1((n_2 + 2)^2 - \ell^2)}$$

The polyad quantum number $P = 2n_1 + n_2$ and the vibrational angular momentum $\ell$ are good quantum numbers for this Hamiltonian. The Hamiltonian matrix factorizes in blocks for each value of $P$ and $\ell$. For practical purposes, we label the eigenstates by a single index $j$. In computations reported in Sect. 5, $\Phi_{\overline{\mathbf{p}},\overline{\mathbf{q}}}(q_1, q_2, \varphi, t = 0)$ was projected on all eigenstates $\Psi_j(q_1, q_2, \varphi)$ with $P \leq 22$ and $|\ell| \leq 10$.

**Appendix C : More theoretical and geometrical aspects**

In this Appendix, we give relations between various sets of dynamical variables used here and in related work [3,5,6] to describe a two-dimensional harmonic oscillator in resonance 1:1 and its perturbations. In $CO_2$ these variables represent bending vibrations (mode 2), in the swing-spring they describe pendular oscillations. We then comment on our treatment of the whole system with three degrees of freedom and compare it to Refs. [3,5,6].

Before we begin, we like to draw attention to the difference in designating resonance systems that can cause some confusion. Near an equilibrium, the usual definition of the resonance condition is given in mechanics (see for example Appendix (7.A) in Ref. [15]) in terms of the ratio of frequencies of the linear approximation (i.e., harmonic frequencies) corresponding to each degree of freedom. It is customary to sort frequencies in ascending order. According to such definition, the system of symmetric stretch vibration (mode 1) and bending vibration (degenerated mode 2) of $CO_2$ has frequency ratio $\omega_2 : \omega_2 : \omega_1$ which is almost exactly 1:1:2 times a common factor. Such plain notation is used in Refs. [3,5,6] and in this Appendix. On the other hand, the traditional molecular notation is 1:2. To complicate matters, many researchers (including Fermi itself) focused exclusively on the $L = 0$ case, which is equivalent to the planar swing-spring system with two degrees of freedom and is also designated 1:2. Furthermore, molecular physicists sometimes require the presence of nonlinear coupling terms in order to qualify the system as resonant: for example they distinguish 1:1 and 2:2 systems. Since the axial symmetry forbids direct coupling of the two components of the bending vibration in $CO_2$, they argue in favor of the 1:2 notation which we used in the main body of the paper.

## C.1 - Coordinates and relations between them

### C.1.1 - Cartesian coordinates

There are two widely used sets of dimensionless Cartesian variables, the initial normal mode coordinates $(q_{2x}, q_{2y})$ and conjugate momenta $(p_{2x}, p_{2y})$ and *rotated* variables $(Q_k, P_k)$ ($k$=2,3), such that

$$(q_{2x}, p_{2x}, q_{2y}, p_{2y}) = \frac{1}{\sqrt{2}}(Q_3 + P_2, P_3 - Q_2, Q_2 + P_3, P_2 - Q_3)$$

as well as related complex Hamiltonian coordinates $z_k = q_k - ip_k$ ($k$=2x,2y) and $Z_k = Q_k - iP_k$ ($k$=2,3), which satisfy $\{z_k, \bar{z}_k\} = \{Z_k, \bar{Z}_k\} = 2i$. The advantage of $(Q_k, P_k)$ is that in these variables both the harmonic oscillator Hamiltonian

$$\begin{aligned} H_0 &= \frac{1}{2}(p_{2x}^2 + q_{2x}^2 + p_{2y}^2 + q_{2y}^2) = \frac{1}{2}(\bar{z}_{2x} z_{2x} + \bar{z}_{2y} z_{2y}) \\ &= \frac{1}{2}(P_2^2 + Q_2^2 + P_3^2 + Q_3^2) = \frac{1}{2}(\bar{Z}_2 Z_2 + \bar{Z}_3 Z_3) \end{aligned} \quad (C.1)$$

and the angular momentum

$$\begin{aligned} L &= q_{2x} p_{2y} - p_{2x} q_{2y} = \frac{i}{2}(\bar{z}_{2x} z_{2y} - z_{2x} \bar{z}_{2y}) \\ &= \frac{1}{2}(P_2^2 + Q_2^2) - \frac{1}{2}(P_3^2 + Q_3^2) = \frac{1}{2}(\bar{Z}_2 Z_2 - \bar{Z}_3 Z_3) \end{aligned} \quad (C.2)$$

have diagonal representation.

### C.1.2 - Polar coordinates

In the presence of axial symmetry, many prefer using polar coordinates and conjugate momenta

$$r = q_2 = \sqrt{q_{2x}^2 + q_{2y}^2}$$
$$\varphi = \tan^{-1}\frac{q_{2y}}{q_{2x}}$$
$$p_r = p_2 = p_{2x} \cos\varphi + p_{2y} \sin\varphi$$
$$p_\varphi = L$$

because in these coordinates the Hamiltonian of an axially symmetric system (such as $CO_2$) does not depend on the cyclic variable $\varphi$, $L$ can be immediately treated as a constant of motion with value $\ell$, and the system can thus be reduced to one degree of freedom with reduced phase space $\mathbf{R}^2$ and coordinates $(q_2, p_2)$ [16]. In particular,

$$H_0 = \frac{1}{2}(p_2^2 + q_2^2 + \frac{L^2}{q_2^2}) \ .$$

The major disadvantage of this very traditional approach is that the singularity of polar coordinates at $q_2 = p_2 = 0$ distorts the geometry of the system. In fact, the true topology of the reduced phase space obtained after reduction of the axial symmetry SO(2) is not that of $\mathbf{R}^2$ but that of a cone with vertex at $q_2 = p_2 = 0$. Polar coordinates are inconvenient in the analysis of monodromy and plane switching because in these studies we are interested in the dynamics near the hyperbolic equilibrium $q_2 = p_2 = 0$ [3,8,17]. In this paper, we do not use polar coordinates except for some intermediate results and for representation of quantum wave functions (which is yet another well established tradition).

### C.1.3 - Action-angle coordinates

There are two varieties of action-angle variables for the two-oscillator in 1:1 resonance. One is simple one-dimensional oscillator action variables $I_k$ and conjugate angles $\phi_k$

$$I_k = \frac{1}{2}(q_{2k}^2 + p_{2k}^2) = \frac{1}{2}\bar{z}_k z_k$$

$$\phi_k = -\tan^{-1}\frac{p_k}{q_k} = \arg z_k$$

where $k=2x, 2y$, for the bending mode 2. Notice that similar variables $(I_1, \phi_1)$ are used in this paper for the nondegenerate stretching mode 1. The other variety is the pair $(I_2, L)$, where $I_2 = H_0$ is the total mode 2 action, and conjugate angles $(\phi_2, \phi_L)$. This latter case comes up naturally when (i) the zero order Hamiltonian, which defines approximate dynamical symmetry of the perturbed system, is $H_0$, and (ii) the system is axially symmetric and $L$ is the first integral. As can be easily seen from the Cartesian definitions in Eqs. (C.1) and (C.2),

$$\phi_2 = \frac{1}{2}(\arg Z_2 + \arg Z_3) = \frac{1}{2}\tan^{-1}\frac{P_3 Q_2 + P_2 Q_3}{P_2 P_3 - Q_2 Q_3} = \frac{1}{2}\tan^{-1}\frac{q_{2x}^2 + q_{2y}^2 - p_{2x}^2 - p_{2y}^2}{2(p_{2x}q_{2x} + p_{2y}q_{2y})}$$

$$\phi_L = \frac{1}{2}(\arg Z_3 - \arg Z_2) = \frac{1}{2}\tan^{-1}\frac{P_3 Q_2 - P_2 Q_3}{P_2 P_3 + Q_2 Q_3} = \frac{1}{2}\tan^{-1}\frac{q_{2y}^2 + p_{2y}^2 - q_{2x}^2 - p_{2x}^2}{2(q_{2x}q_{2y} + p_{2x}p_{2y})}$$

At this point it is instructive to notice that $\phi_2$ is *not* $\zeta = (\phi_{2x} + \phi_{2y})/2$ as one could naively assume looking at Eq. (C.1) (check: $\{\zeta, H_0\} = 1$ but $\{\zeta, L\} \neq 0$ and so $\zeta$ is not the required angle variable). For similar reasons, $\phi_L \neq \varphi$. Instead one verifies easily that

$$\rho = \phi_L + \frac{\pi}{4} = \varphi - \frac{1}{2}\tan^{-1}\frac{Lp_2 q_2}{L^2 - q_2^2 H_0} \ .$$

The angle $\rho$ plays the central role in the description of plane switching because it gives the instantaneous position of the plane (or more precisely, of the major axis of the ellipse which the trajectory traces in the $(x,y)$ plane) [3]. Notice that in Ref. [3] this angle is denoted as $\theta$.

*C.1.4 - Angular momentum analogy for the 1:1 oscillator system*

In physical applications, the $S^1$ dynamical symmetry of the 1:1 resonance is often approximate and the original Hamiltonian should be first normalized. If the original system has other degrees of freedom which are not in resonance, they can be averaged out at the same time. The resulting normal form Hamiltonian H describes such other degrees effectively and is $S^1$ symmetric, i.e., $H_0$ becomes a constant of motion and $\{H, H_0\} = 0$. In practical interpretations of experimental data one often skips normalization and introduces phenomenological model systems with exact $S^1$ symmetry and with model effective Hamiltonians whose parameters are adjusted to reproduce the data. In molecules, vibrational states with the same $H_0$ are often called polyads, hence the terminology "polyad Hamiltonian", "polyad quantum number", etc. Instructive examples of such polyads can be found in the literature on the triatomic molecules $O_3$ and $H_3^+$.

To represent $H$, we notice that the ring of *all* invariants of the $S^1$ action of the dynamical symmetry of the 1:1 oscillator system is generated multiplicatively by four quadratic invariants of the general form $\bar{z}_k z_m$. They can be chosen as

$$j = \frac{1}{4}(\bar{z}_{2x} z_{2x} + \bar{z}_{2y} z_{2y}) = \frac{H_0}{2}$$
$$j_1 = \frac{1}{4}(\bar{z}_{2x} z_{2x} - \bar{z}_{2y} z_{2y})$$
$$j_2 = \frac{i}{4}(\bar{z}_{2x} z_{2y} - \bar{z}_{2y} z_{2x}) = \frac{L}{2} \qquad (C.3)$$
$$j_3 = \frac{1}{4}(\bar{z}_{2x} z_{2y} + \bar{z}_{2y} z_{2x})$$

and are subject to the sole algebraic relation of degree 2

$$j_1^2 + j_2^2 + j_3^2 - j^2 = 0 \ . \qquad (C.4)$$

The widely used analog of this construction in quantum mechanics is the boson representation of angular momentum operators which is introduced by Schwinger [18] and which is based on the isomorphism of the algebras su(2) and so(3).

It can be seen that (as any $S^1$ invariant) the classical polyad Hamiltonian $H$ is a function of $(j_1, j_2, j_3)$ and $j$. Since $j$ is a constant of motion, we fix $j$ and consider it as a dynamical parameter thus reducing the original 1:1 oscillator system to one degree of freedom. The reduced Hamiltonian $H_j(j_1, j_2, j_3)$ is defined on the reduced (or polyad) phase space $P_j$. From Eq. (C.4) we see that for any $j > 0$ this space is a sphere $\mathbf{S}^2$ and indeed it often is called "polyad sphere". At this point we can appreciate the advantage over any approach with one fixed system of canonical coordinates as any such system for $\mathbf{S}^2$ would have singularities! In fact, to work on $\mathbf{S}^2$ we should use (at least) two symplectic charts - or run into trouble. Thus in the presence of axial symmetry (which acts on $(j_1, j_2, j_3)$ as rotation about axis $j_2$), one is often tempted to use $L = 2j_2$, known as "vibrational" angular momentum, and the corresponding conjugate angle

$$\frac{1}{2}\phi_{j_2} = \frac{1}{2}\tan^{-1}\frac{j_1}{j_2} = \phi_L$$

which follows for the 3-vector $\mathbf{j} = (j_1, j_2, j_3)$ and is, of course, the same as in Sec. C.1.3. It can be seen that coordinates $(L, \phi_L)$ with $|L| < 2j$ and $0 \leq \phi_L \leq 2\pi$ define a cylindrical chart of $\mathbf{S}^2$ with obvious problems at the poles $L = \pm 2j$. Consequently, $(L, \phi_L)$ should be used with appropriate caution.

Finally notice that canonical polyad coordinates are simply unnecessary. Their purpose in typical polyad studies seems to be the derivation of Hamilton equations of motion for the reduced system. However, for any given classical polyad Hamiltonian $H_j = H_j(j_1, j_2, j_3)$, the dynamics of the reduced system, i.e., the internal polyad dynamics, is described by the Euler-Poisson equations $dj_k / dt = \{j_k, H\}$ with $k=1,2,3$. These are similar to the equations for rotating nonrigid bodies and can be found easily once we compute the Poisson algebra generated by $(j_1, j_2, j_3)$. Not surprisingly, the latter is an su(2)~so(3) algebra with standard structure

|       | $j_2$ | $j_3$  |
|-------|-------|--------|
| $j_1$ | $j_3$ | $-j_2$ |
| $j_2$ |       | $j_1$  |

$$\tag{C.5}$$

and Casimir $j$.

**C.2 - Reduced $CO_2$ Hamiltonian and reduced phase space**

The plane switching phenomenon in $CO_2$ and the swing-spring is studied after reducing the original system with three degrees of freedom both with respect to the axial symmetry SO(2) of rotations of plane $(q_{2x}, q_{2y})$ already mentioned above, and the dynamical symmetry $S^1$ defined by the flow of the zero order Hamiltonian

$$I = 2I_1 + I_2 = J + I_2 = p_1^2 + q_1^2 + \frac{1}{2}(p_{2x}^2 + p_{2y}^2 + q_{2x}^2 + q_{2y}^2)$$

which represents a three-dimensional harmonic oscillator with frequencies in 1:1:2 resonance. The original $CO_2$ Hamiltonian is, of course, not strictly $S^1$ invariant. It is made so by normalization (the particular method used in Ref. [5] and in this work is CPT [11]) and truncation at the desired order.

*C.2.1 - Geometry of singular reduction*

The truncated normal form $H$ in Eq. (2.3) Poisson commutes both with $I$ and (since normalization preserves the axial symmetry) with $L$. In other words, $I$ and $L$ are first integrals of the system with Hamiltonian $H$. Reduction of the respective degrees of freedom means reduction of the combined action of the dynamical 1:1:2 symmetry $S^1$ and the axial symmetry SO(2). As a result we obtain a system with one degree of freedom and two-dimensional reduced phase space $P_{I,L}$. There is a one-to-one correspondence between points of spaces $P_{I,L}$ and orbits of the $S^1 \times SO(2)$ action.

As we show later in Sec. C.2.3, the space $P_{I,L}$ does not always have the same topology : for $|L|=I$ it is just a point, it is diffeomorphic to a sphere $\mathbf{S}^2$ for all $L \neq 0$ and $|L|<I$ (see Fig. 7, right), while $P_{I,0}$ is a *singular sphere* with one conical point for maximum $J=I$ (see Fig. 7, left). The geometric reason for the singularity of $P_{I,0}$ is fundamental. The 1:1:2 action $S^1$ on the original phase space $\mathbf{R}^6_{q,p}$ of the Fermi system is not free because its circular orbits defined by $p_1^2+q_1^2=2I$ in the plane $\sigma^* = \{p_{2x}=p_{2y}=q_{2x}=q_{2y}=0\}$ are, obviously, two times shorter than all others (for $CO_2$ these are trajectories of the symmetric stretch normal mode 1). The SO(2) action is not free either because points in $\sigma^*$ (and only these points) are its fixed points. It follows that for fixed $I>0$ and $L=0$, the combined $S^1 \times$ SO(2) action has *one* nonregular circular orbit $\mathbf{S}^1_I \subset \sigma^*$ with $J=I$, while all other orbits are regular 2-tori. Reduction of the $S^1 \times$ SO(2) symmetry sends these orbits to points of the reduced spaces $P_{I,L}$. Every point of $P_{I,L}$ with $L \neq 0$ represents a particular regular $\mathbf{T}^2$ orbit, while $P_{I,0}$ has one isolated point which represents the singular circle. Such point on $P_{I,0}$ must necessarily be singular. The situation described above is characteristic of reduction of non-free symmetry actions or singular reduction pioneered by Cushman [8,17] as a generalization of regular reduction of Lie symmetries [19]. Singular reduction is particularly common in applications such as systems with resonances other than 1:...:1.

*C.2.2 - Reduction using action-angle variables*

To represent the normalized Hamiltonian $H$ in this paper, we use action variables $(I,L,J)$, such that $|L| \leq I$ and $0 \leq J \leq I-|L|$. Of the three corresponding conjugate angles, $\phi_I$ and $\phi_L$ are, obviously, cyclic variables not present in $H = H(J, \phi_J, I, L)$ where $\phi_J$ is conjugate to $J$ (see Eq. (2.2)). Replacing $I$ and $L$ for their values, we obtain the *reduced* Hamiltonian $H_{I,L}(J, \phi_J)$.

Coordinates $(J, \phi_J)$ are "polar coordinates in disguise" [17]: as illustrated in Fig. 7, they give a cylindrical chart $\mathcal{P}_{I,L} = [0, I - |L|] \otimes \mathbf{S}^1$ of the actual reduced phase space $P_{I,L}$. The situation is similar to that discussed in Sec. C.1.4 albeit now $P_{I,L}$ has not always the $\mathbf{S}^2$ topology and can itself be singular. Worst of all, the singularity of $P_{I,0}$ is masked by the singularity of the chart $P_{I,L} \to \mathcal{P}_{I,L}$ and it remained for this reason unrecognized in a number of studies of the polyads of the 1:2 Fermi systems.

One might argue that coordinates $(J, \phi_J)$ are inappropriate for studying monodromy and plane switching in our system because in such study we are interested in the dynamics near the "combined" singularity of these coordinates at $J = I$ and the underlying phase space $P_{I,0}$ and should have discontinuities in the solution for $\phi_J(t)$. However, following the derivation of Eqs. (2.4) and (2.10), one realizes that $\phi_J(t)$ is of no value to our analysis and that we have circumvented this difficulty by excluding $\phi_J(t)$ completely from our consideration. In particular Eq. (2.10) is obtained by combining $d\rho/dt = \partial H/\partial L$ and $H = E$ in order to eliminate the Fermi term with $\cos(2\phi_J)$.

*C.2.3 - Poisson reduction*

We now briefly survey the way to reduce the symmetries of the 1:1:2 system while preserving its geometry [3,5,6] in order to show that our Eqs. (2.4) and (2.10) are indeed direct higher order analogs of the ones studied previously for the resonant swing-spring [3]. The reduced system can be described fully using three dynamical variables

$$R = \frac{1}{4}(\bar{z}_2 z_2 + \bar{z}_3 z_3) = \frac{1}{2}(I - J)$$

$$S = \frac{1}{4}(\bar{z}_1 z_2^2 + z_1 \bar{z}_2^2 + \bar{z}_1 z_3^2 + z_1 \bar{z}_3^2)$$ (C.6)

$$T = \frac{i}{4}(\bar{z}_1 z_2^2 - z_1 \bar{z}_2^2 + \bar{z}_1 z_3^2 - z_1 \bar{z}_3^2)$$

which are all invariants of the combined action of the 1:1:2 oscillator symmetry $S^1$ and axial symmetry SO(2) and obey

$$2\Phi_{I,L} = T^2 + S^2 - (4R^2 - L^2)(I - 2R) = 0$$ (C.7)

and inequalities $0 \leq |L|/2 \leq R \leq |I|/2$. We should also notice that the reduced system inherits additional finite symmetry properties from the original system, notably the invariance with respect to $T \to -T$ and $L \to -L$. Taking these extra discrete symmetries and Eq. (C.7) into account, we can show that the reduced Hamiltonian $\mathcal{H}_{I,L}$ is an arbitrary polynomial in just *two* invariants $R$ and $S$ and that it includes only even degrees in $L$. The trio of functions $(R, S, T)$ generate the Poisson algebra of the reduced system. By a direct computation we find the structure of this algebra

|   | $S$ | $T$ |
|---|-----|-----|
| $R$ | $T$ | $-S$ |
| $S$ |   | $12R^2 - 4IR - L^2$ |

Note that function $\Phi_{I,L}$ in Eq. (C.7) is the Casimir and

$$\{\chi_a, \chi_b\} = \varepsilon_{abc} \frac{\partial \Phi_{I,L}}{\partial \chi_c}$$ (C.8)

for $\chi = (\chi_1, \chi_2, \chi_3) = (R, S, T)$. We can now define the Poisson structure on $P_{I,L}$ and find equations of motion for dynamical variables $(R, S, T)$ of the reduced system. For the second-order Hamiltonian of degree 4 in $(p_k, q_k)$

$$\mathcal{H} = aS - b(I)R + cR^2 - h_0(I, L^2) = h \tag{C.9}$$

we have

$$\frac{dR}{dt} = \{R, \mathcal{H}\} = aT$$

$$\frac{dS}{dt} = \{S, \mathcal{H}\} = (b - 2cR)T$$

$$\frac{dT}{dt} = \{T, \mathcal{H}\} = -12aR^2 + 4aIR - (b - 2cR)S + aL^2$$

Solving Eq. (C.7) for $T^2(S, R, I, L)$ and Eq. (C.9) for $S(R, h)$ and substituting in $(dR/dt)^2$ gives the equivalent of Eq. (2.4)

$$\left(\frac{dR}{dt}\right)^2 = -c^2 R^4 + (2bc - 8a^2)R^3 + (4Ia^2 + 2hc - b^2)R^2 + 2(L^2 a^2 - hb)R - h^2 - IL^2 a^2$$

(C.10)

Similarly, we can obtain the second order o.d.e. from $d^2 R/dt^2 = a\, dT/dt$ which corresponds to Eq. (59) in Ref. [3]. Notice that the reduced Hamiltonian remains linear in $S$ up to order 3 and that the same approach can be used to find equations of motions at this order.

When $c = 0$, solution $R(t)$ of Eq. (C.10) can be expressed in terms of the Weierstrass elliptic function $\wp$ [20]

$$R(t) = c_0 + c_1 \wp(t + it_0; g_2, g_3) \tag{C.11}$$

This kind of solution (for $b = 0$) was used previously in the studies of the model 1:1:2 resonant swing-spring system [1-4]. Solution for the case $c \neq 0$ is given in Ref. [10] in terms of Jacobi elliptic function sn(t) and is used in this paper. This solution is a rational function of $\wp$. In our case of $c \ll a$ it can be written as a sum of solution in Eq. (C.11) (with modified parameters) and a small correction which is a rational function of $\wp$ and which vanishes in the limit $c \to 0$.

Finally, we derive equation for $d\rho/dt$ which is easier to obtain in the action-angle variables as $\partial H_{I,L}(J,\phi_J)/\partial L$. Here we notice that $\phi_L$ Poisson commutes with $R$ by definition. Furthermore,

$$0 = \{\{\phi_L,S\},L\} + \{\{S,L\},\phi_L\} + \{\{L,\phi_L\},S\} = \{\{\phi_L,S\},L\} + \{0,\phi_L\} - \{1,S\} = \{\{\phi_L,S\},L\}$$

and it is also clear that $\{\{\phi_L,S\},I\} = 0$. Therefore $d\rho/dt$ is a function on the reduced phase space and can be expressed in terms of $(R,S,T)$. Direct computation using the $(p_k,q_k)$ definition of $\phi_L$ in Sec. C.1.3 gives

$$\frac{d\rho}{dt} = \{\rho,\mathcal{H}\} = a\{\phi_L,S\} - \frac{\partial h_0}{\partial L} = a\frac{LS}{L^2 - 4R^2} - \frac{\partial h_0}{\partial L}$$

Replacing $S$ for solution $S(R,h)$ of Eq. (C.9) we obtain the equivalent of Eq. (2.10).

# FIGURE CAPTIONS

**Figure 1** (color online) : Variation of the classical angular frequency $\omega^*$ (in cm$^{-1}$) as a function of energy $E$ (in cm$^{-1}$) for $CO_2$ at $I = 13.5$ and various values of $L$.

**Figure 2** (color online) : Variation of the angle of plane switching $\Delta\rho - 2x_{LL}LT$ (in radians) as a function of energy $E$ (in cm$^{-1}$) for $CO_2$ at $I = 13.5$ and various values of $L$.

**Figure 3** (color online) : Variation of the angle $\rho$ (in radians) as a function of time $t$ (in fs) for $CO_2$ at $I = 13.5$ and $L = 0.1$ (top plot), $L = 2$ (middle plot) and $L = 6$ (bottom plot). For each value of $L$, the variation of $\rho$ is shown for three trajectories starting at maximum $J = \alpha$ and with respective energies $E = E_{I-L} + 0.1$ cm$^{-1}$, $E = E_{I-L} + 40$ cm$^{-1}$, and $E = E_{I-L} + 400$ cm$^{-1}$.

**Figure 4** (color online) : Projection on the $(q_{2x}, q_{2y})$ plane of trajectories with $I = 13.5$, $L = 0.1$ (top plot), $L = 2$ (middle plot) or $L = 6$ (bottom plot), and $E = E_{I-L} + 0.1$ cm$^{-1}$. "0" indicates the starting point of the trajectory, "1" the direction after the first plane switch and "2" the direction after the second plane switch. The trajectories with $L = 0.1$, $L = 2$ and $L = 6$ were integrated for 750, 385 and 260 fs, respectively.

**Figure 5** (color online) : Projection on the $(q_{2x}, q_{2y})$ plane, at various times $t$, of the swarm of 10000 trajectories of $CO_2$ with initial gaussian distribution centered on a trajectory with $I = 13.5$, $L = 2$ and $E = E_{I-L} + 0.1$ cm$^{-1}$. The position of this trajectory is marked for each time *t* by an empty circle (compare with middle plot of Fig. 4). The time *t* (in fs) is indicated in the upper left corner of each vignette.

**Figure 6** (color online) : Contour plots of the probability density $P(q_{2x}, q_{2y}, t)$ at different times *t* (*t* is indicated in the upper left corner of each vignette). At time $t = 0$, the quantum wave packet is centered on the same trajectory with $I = 13.5$, $L = 2$ and $E = E_{I-L} + 0.1$ cm$^{-1}$ as in the case of the swarm of classical trajectories of Fig. 5. The position of this trajectory is marked for each time *t* by an empty circle.

**Figure 7** : Singular (left) and regular (right) reduced phase spaces $P_{I,L}$ and their cylindrical charts $\mathcal{P}_{I,L}$.

**Figure 1**

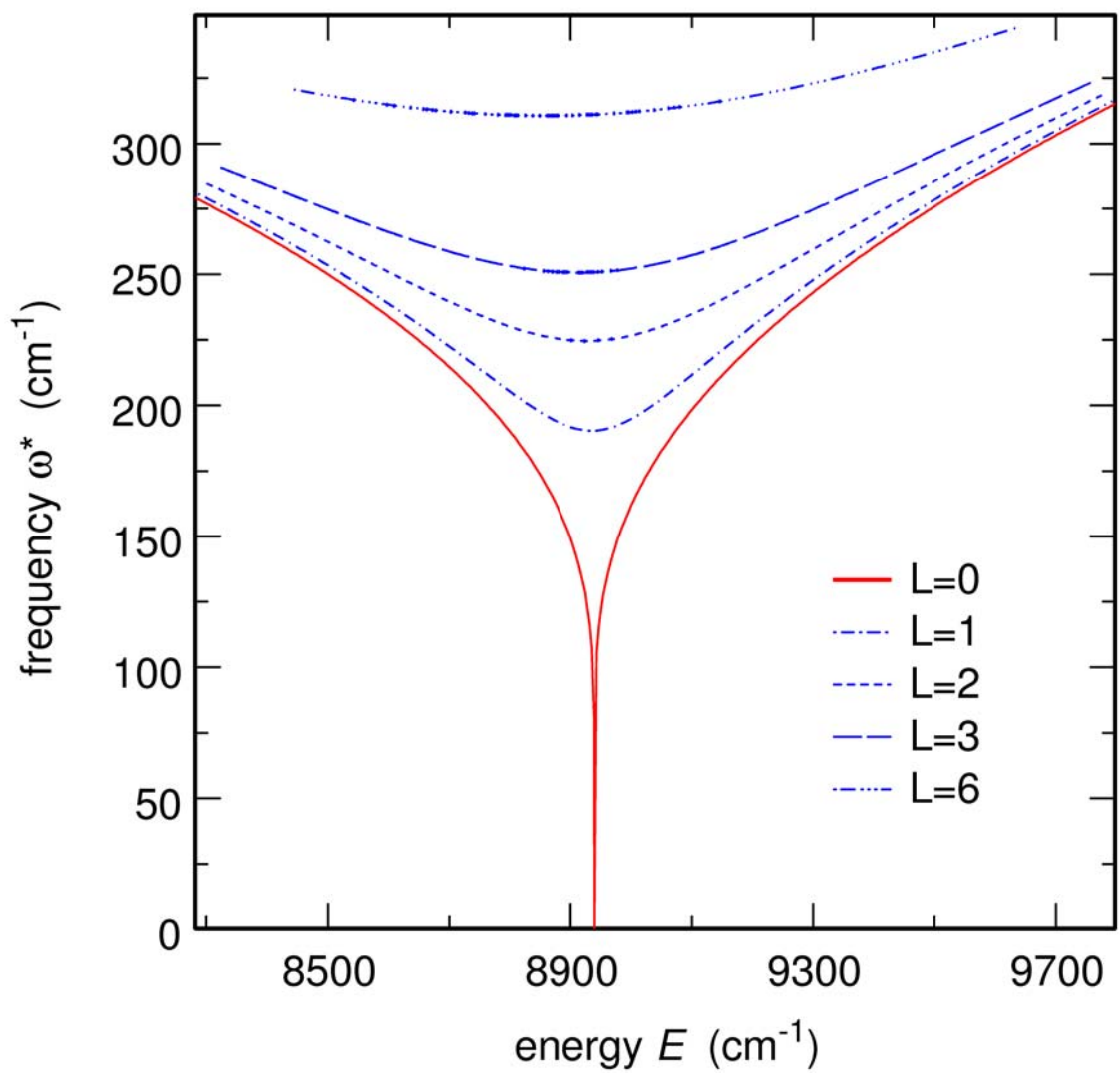

**Figure 2**

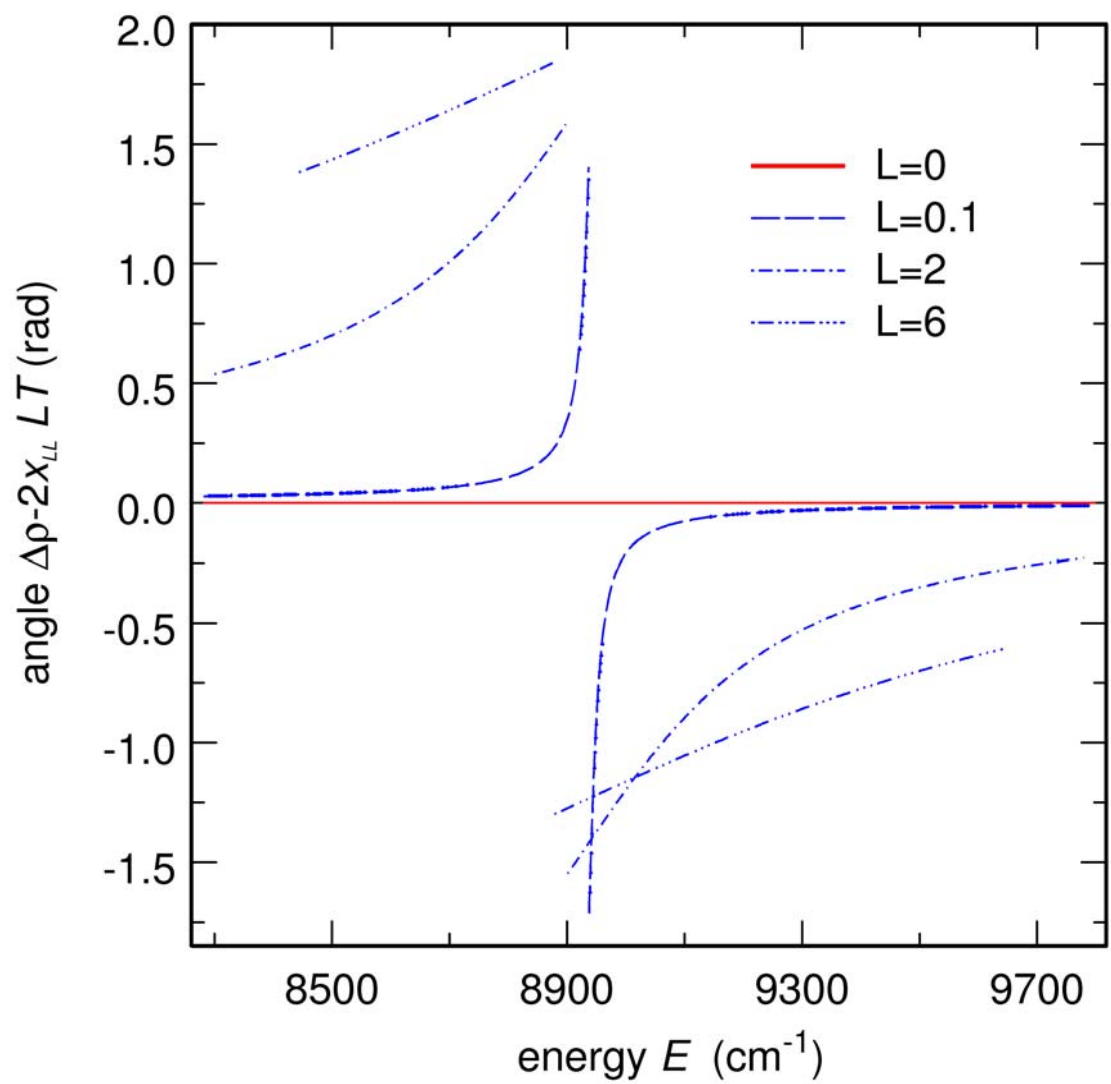

**Figure 3**

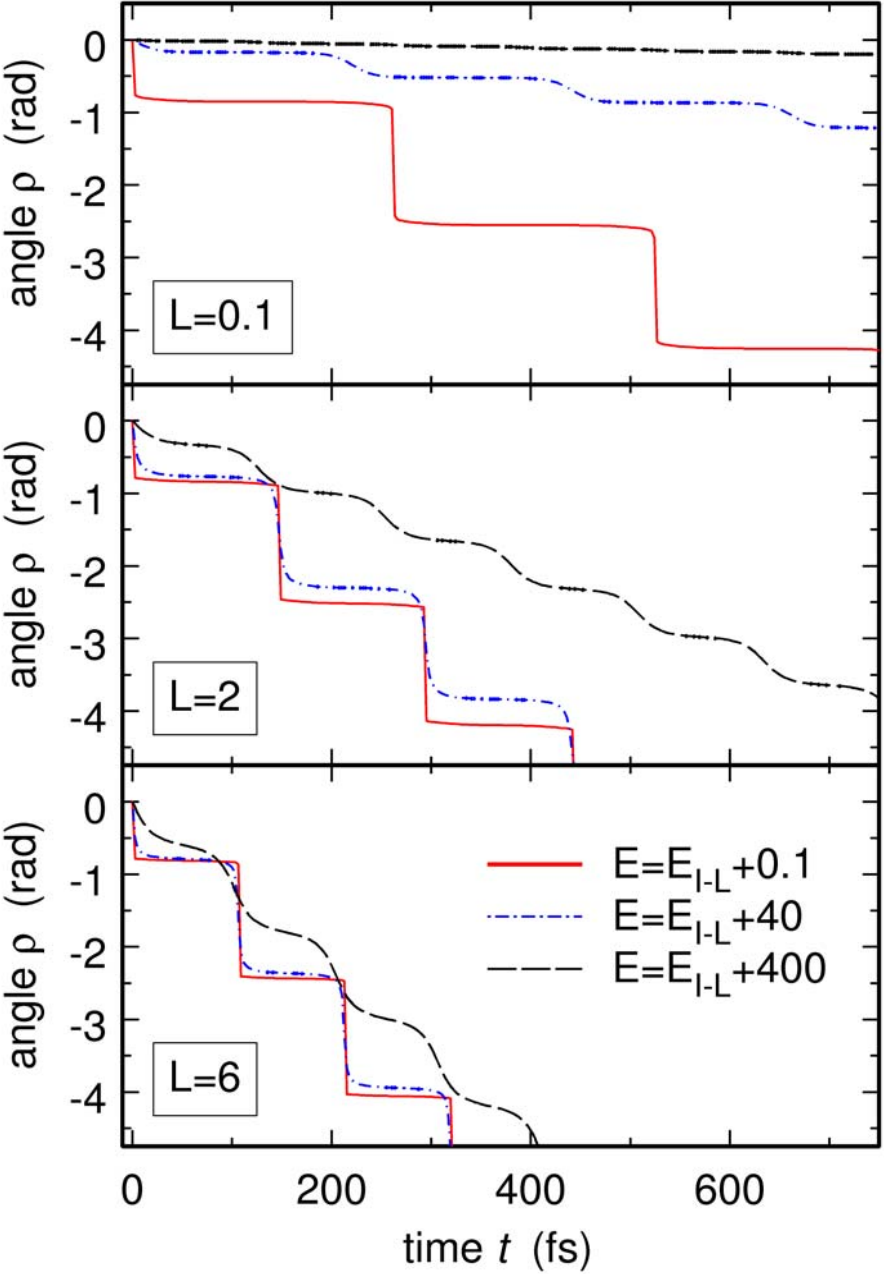

**Figure 4**

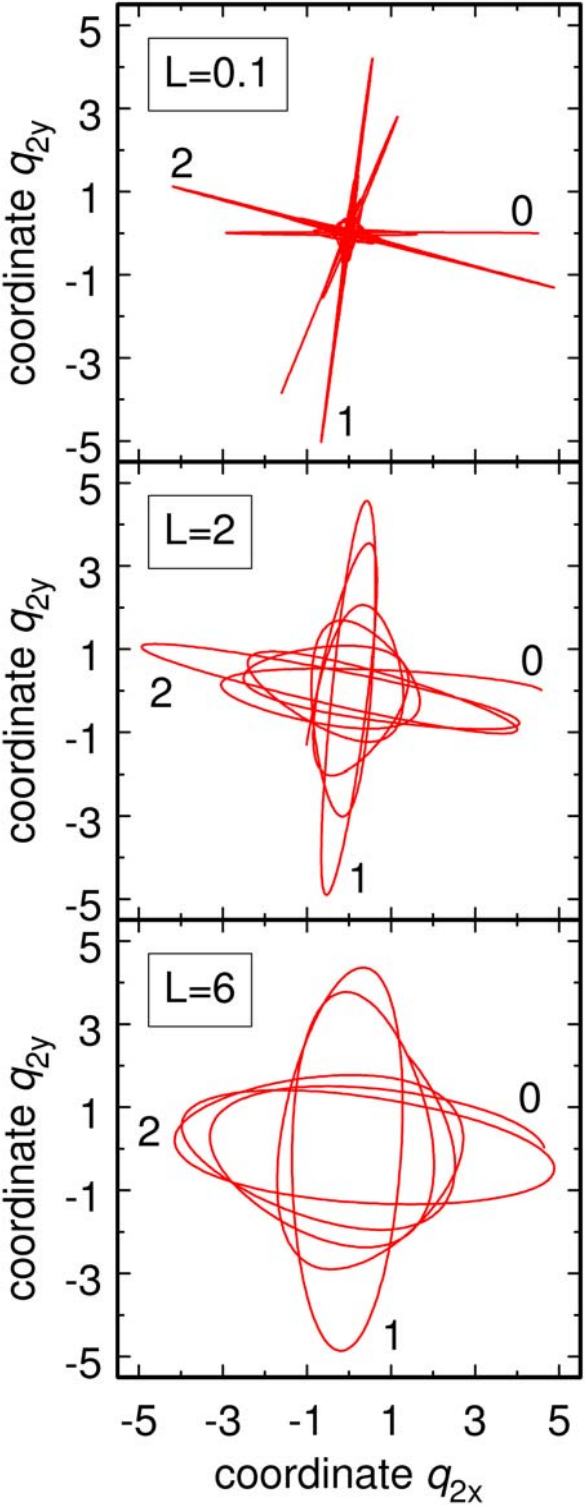

**Figure 5**

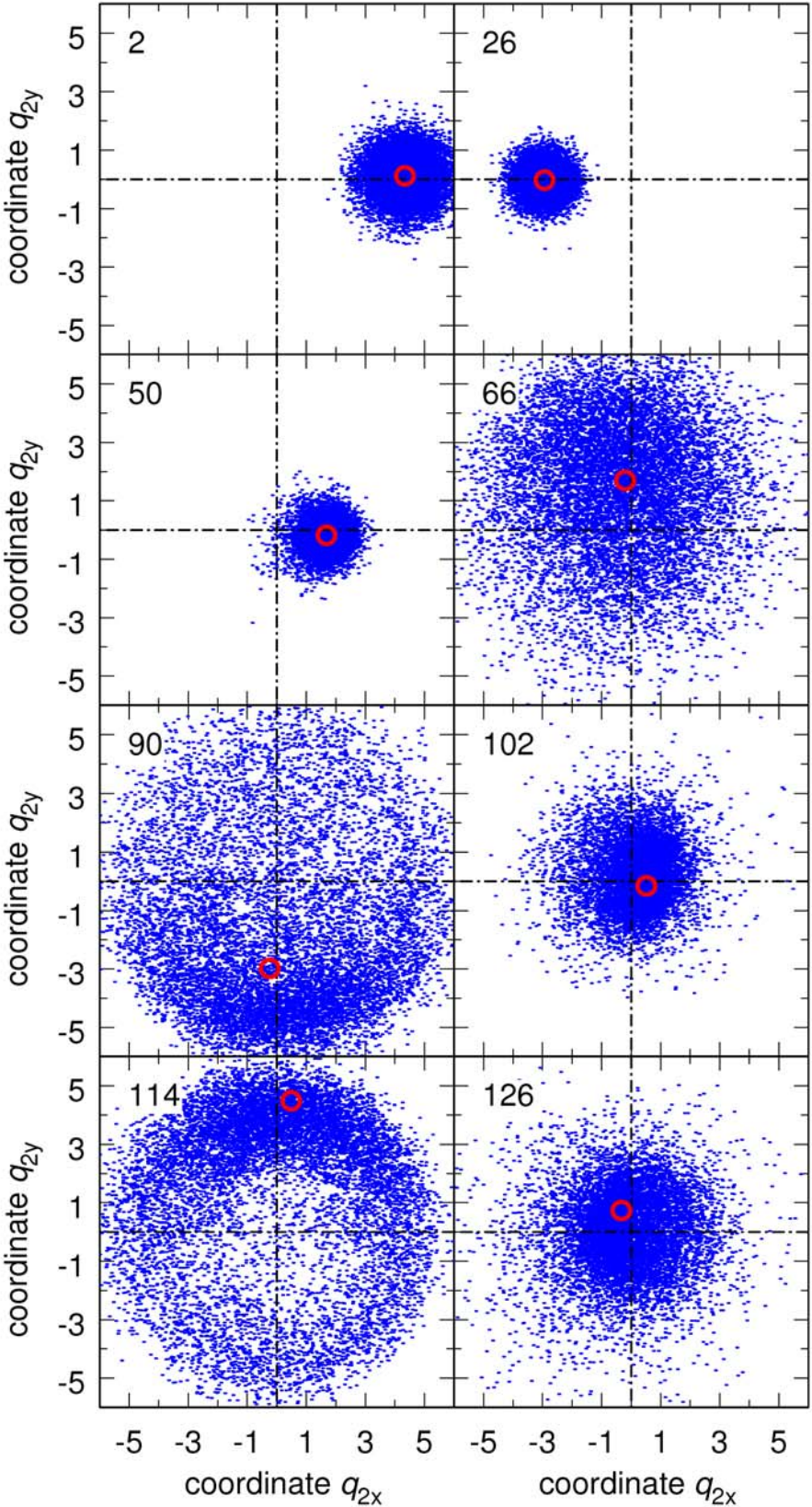

**Figure 6**

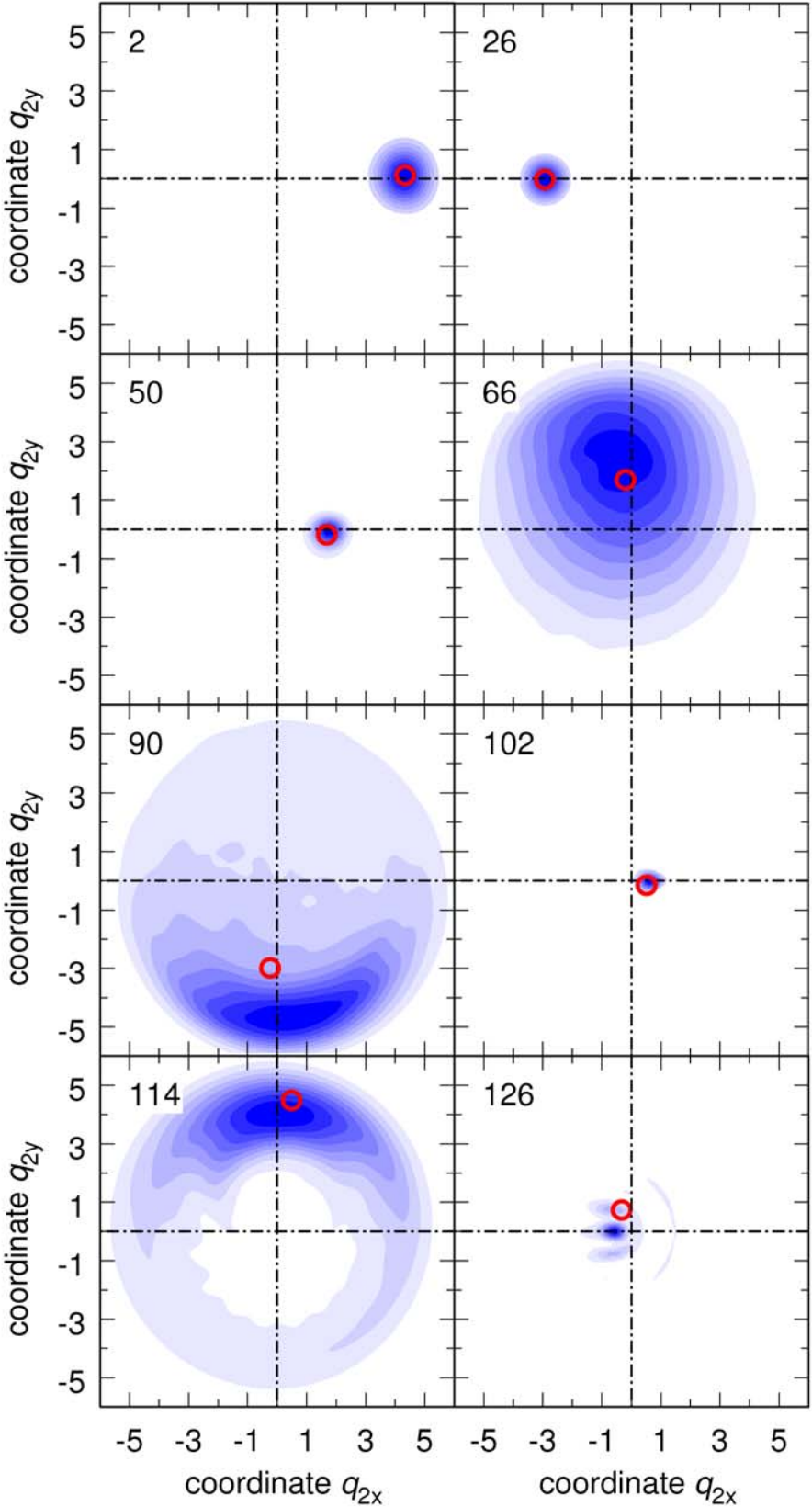

**Figure 7**

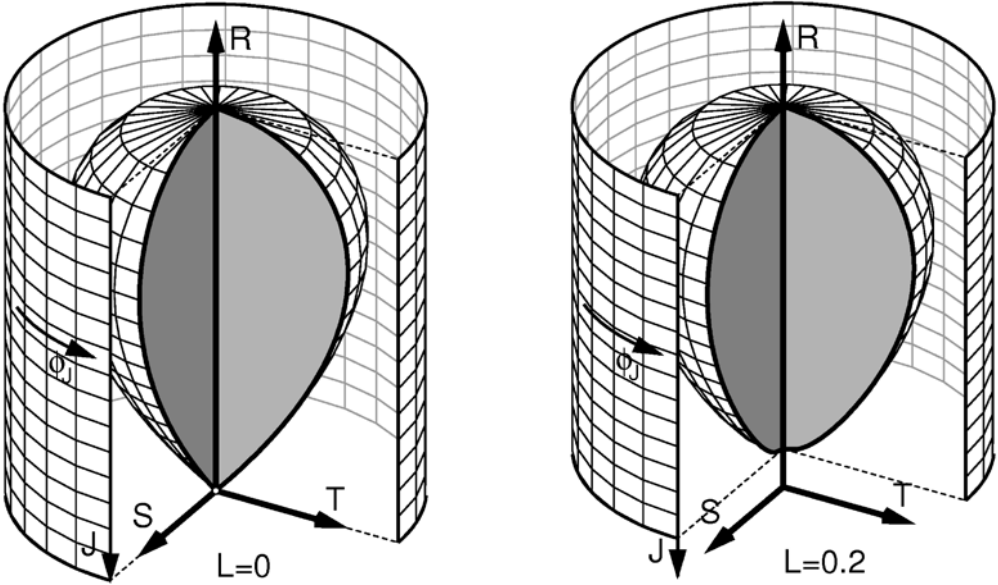